\documentclass[preprint,preprintnumbers, prd, floatfix, superscriptaddress,nofootinbib] {revtex4-1}
\usepackage{graphicx}
\usepackage{epsfig}
\usepackage{bm}
\usepackage{amssymb}
\usepackage{amsmath}
\usepackage{dcolumn}
\usepackage{cancel}
\usepackage[colorlinks]{hyperref}
\usepackage{epstopdf}
\hypersetup{
     breaklinks=true,
    pdfstartview={FitH},    colorlinks=true,       
    linkcolor=blue,          
    citecolor=red,        
    filecolor=magenta,      
    urlcolor=blue,           
    anchorcolor=green,      
    linktocpage=true
}

\begin{document}

\title{Cosmological constraints on  $\Lambda(\alpha)$CDM models
with time-varying fine structure constant}

\author{Jin-Jun Zhang}
\email{zhangjinjun@sxnu.edu.cn}
\affiliation{School of Physics and Information Engineering, Shanxi Normal University, Linfen 041004, China}
\author{Lu Yin}
\email{yinlu@gapp.nthu.edu.tw}
\affiliation{Department of Physics, National Tsing Hua University, Hsinchu, Taiwan 300}
\author{Chao-Qiang Geng}
\email{geng@phys.nthu.edu.tw}
\affiliation{Department of Physics, National Tsing Hua University, Hsinchu, Taiwan 300}
 \affiliation{National Center for Theoretical Sciences, Hsinchu,
Taiwan 300}
\affiliation{School of Physics and Information Engineering, Shanxi Normal University, Linfen 041004, China}

\begin{abstract}
We  study the $\Lambda(\alpha)$CDM models with $\Lambda(\alpha)$ being a function of
the time-varying fine structure constant $\alpha$.
We give a close look at the constraints on two specific $\Lambda(\alpha)$CDM models
with one and two model parameters, respectively,
based on  the cosmological observational measurements along with  313 data points
for the time-varying $\alpha$.
We find that the model parameters are constrained to be around $10^{-4}$, which are similar to the results
discussed previously  but  more accurately.

\end{abstract}

\maketitle

\section{Introduction}
The cosmological constant ($\Lambda$)  was first introduced to the general theory of relativity
by Einstein~\cite{1EINSTEN} more than one hundred years ago~\cite{100yrs}. Nowadays, it is contained in
the standard model of cosmology: $\Lambda$ cold dark matter ($\Lambda$CDM), which is the simplest way to act as
dark energy~\cite{DE}
to explain the current accelerated expanding universe discovered in 1998~\cite{Riess:1998cb, Perlmutter:1998np}.
However,  in the $\Lambda$CDM model there is a well known cosmological constant problem, related to
the two  theoretical difficulties of
 ``coincidence''~\cite{Ostriker:1995rn, ArkaniHamed:2000tc} and  ``fine-tuning''~\cite{Weinberg:1988cp, WBook}.
 Note that the fine-tuning one is about the question of ``why the non-zero cosmological constant is so tiny,''
which was known even before the proposal of dark energy in 1998~\cite{Weinberg:1988cp, WBook}.
Although it is believed that the  $\Lambda$ problem
can  be ultimately solved  only in
a unified theory of quantum gravity and the standard model of electroweak and strong interactions in particle physics,
there have been  many  attempts trying to understand this problem
in resent years~\cite{100yrs,Weinberg:1988cp,Review1}.
 In particular, the axiomatic approach~\cite{9:Beck:2008rd} is one of the
most interesting ideas, in which  $\Lambda$ is  derived from four axioms~\cite{11,Boehmer:2005sm,Wesson:2003qn,Boehmer:2006fd},
in close analogy to the Khinchin axioms at the information theory~\cite{15,16,17,18}.

From the four natural and simple axioms, the explicit form of the cosmological constant
is given by~\cite{9:Beck:2008rd}
\begin{equation}
\label{eq:action}
\Lambda=\frac{G^2}{\hbar^4}\left(\frac{m_e}{\alpha}\right)^6 .
\end{equation}
where $G$ is the gravitational constant, $\hbar$ is the reduced Planck constant,
 $m_e$ is the electron mass, and $\alpha$ is the fine structure constant. Note that  the relation  in Eq.~(\ref{eq:action}) has also been independently given in Ref.~\cite{Boehmer:2005sm}.\footnote{For a review on the relation between $\Lambda$ and $\alpha$
 in Eq.~(\ref{eq:action}), see Ref.~\cite{Wei:2016moy}.}
 In 1998,  along with the discovery of the accelerated expansion universe, an evidence of the time variation  of $\alpha$ was
found~\cite{Webb:1998cq,Webb:2000mn},  namely ${\Delta \alpha}/{\alpha}\equiv ({\alpha-\alpha_0})/{\alpha_0}$ is non-zero with
$\alpha_0$ being the present value of  $\alpha$.
It was claimed that $\alpha$ is not only a time varying parameter but a spatially varying one~\cite{King:2012id,Webb:2010hc}.
As a result, in terms of Eq.~(\ref{eq:action}) with  $\Lambda \propto \alpha^{-6}$, the cosmological constant term should
be  time and  space-dependent too~\cite{Uzan:2010pm,Webb:1998cq,Wei:2009xg,Wei:2011jw,Terazawa:2012fa,60}.
In this case, it might be responsible for the possible anisotropy in the accelerated
expansion of the universe.

Recently, a time-varying fine structure constant $\alpha$ has been extensively discussed
in the literature~\cite{Uzan:2010pm,Wei:2009xg,Wei:2011jw,60,Terazawa:2012fa}.
In this scenario, $\alpha$ is only time-dependent  and
increasing with time~\cite{Webb:1998cq,Webb:2000mn,Murphy:2000pz}.
In this paper, unlike the general  $\Lambda(t)$ models without explicit forms, we take those models with $\Lambda(\alpha)  \propto \alpha^{-6}$
in Eq.~(\ref{eq:action}) to study the  time-varying  effects.

In our numerical calculations, we use  the {\bf CAMB}~\cite{Lewis:1999bs} and {\bf CosmoMC}~\cite{Lewis:2002ah} packages with the Markov chain Monte Carlo ({\bf MCMC}) method  to give a close look at the models by including
  313 data points
of ${\Delta \alpha}/{\alpha}$~\cite{King:2012id,30,31,32,Wilczynska:2015una} in {\bf CosmoMC}.
Comparing with  the previous study in Ref.~\cite{Wei:2016moy},
our analysis  starts with a different method of the projection and  a variety of  the observational datasets
together with adding 20 new data points~\cite{Wilczynska:2015una} for ${\Delta \alpha}/{\alpha}$,
resulting in a more accurate outcome.

This paper is organized as follows. In Sec.~\ref{sec:model}, we introduce the varying cosmological constant $\Lambda(\alpha)  \propto \alpha^{-6}$ and derive the
evolution equations for pressureless matter  in the linear perturbation theory. In Sec.~\ref{sec:observation},
we perform the numerical calculations to obtain the observational constraints on the model
parameters  as well as cosmological observables based on the datasets. Our conclusions are
given in Sec.~\ref{sec:CONCLUSIONS}.

\section{Varying cosmological constant models}
\label{sec:model}
\subsection{Formalism}

We consider a spatially flat Friedmann-Robertson-Walker (FRW) universe  with the metric~\cite{Cai:2010hk}
\begin{eqnarray}
\label{eq:fr1}
ds^{2}= -dt^{2}+a^{2}(t) (dr^2+r^2d{\Omega})\
\end{eqnarray}
containing only dark or vacuum energy and  pressureless matter
with the Friedmann equations, given by
\begin{eqnarray}
\label{eq:friedmann1}
&& H^{2}=\frac{8\pi G}{3}(\rho_m  +\rho_{\Lambda}) \,, \\
\label{eq:friedmann2}
&& \dot{H}=- 4\pi G(\rho_m  +\rho_{\Lambda} + P_m + P_{\Lambda}), \,
\end{eqnarray}
where $H=da/(adt)$ is the Hubble parameter
with  $a$  the scale factor,  $\rho_m$ is the energy density of pressureless matter,
$\rho_\Lambda=c^4\Lambda/(8\pi G)$ is the  energy density of dark  energy,
 and $P_{m (\Lambda)}$ is the pressure of pressureless matter (dark energy).
We will describe the varying cosmological constant scenarios in terms of $\rho_\Lambda$  instead of
$\Lambda$.
In the models,  the equation-of-state (EoS) of dark energy (pressureless matter) is given by
$w_{\Lambda(m)}={P_{\Lambda(m)}}/{\rho_{\Lambda(m)}}= -1(0)$.

In this study,  we assume that only the fine structure ``constant" $\alpha = e^2/(\hbar c)$ is varying in time due to
the change of the electric charge $e$, whereas the other fundamental constants $\hbar, G, c$ and $m_e$ are true constants.
Consequently,
$\rho_\Lambda, \Lambda(\alpha) $ and $ \alpha$ are related by $\rho_\Lambda \propto \Lambda(\alpha) \propto \alpha^{-6}$,
which leads to
\begin{equation}
\label{eq:ratio}
\frac{\dot{\rho}_\Lambda}{\rho_\Lambda}=-6\,\frac{\dot{\alpha}}{\alpha}\,.
\end{equation}
 The vacuum energy interacts with  pressureless matter by exchanging energy between them with  the continuity equations, written as
\begin{eqnarray}
\label{eq:eos1}
&& \dot{\rho}_m+3 H\rho_m = Q \,, \\
\label{eq:eos2}
&& \dot{\rho}_\Lambda = - Q \,,
\end{eqnarray}
where the coupling term  $Q =   6 {\rho}_\Lambda \dot{\alpha}/\alpha\ne 0$
from Eq.~(\ref{eq:ratio}).
 The total energy conservation equation is given by
 $\dot{\rho}_{tot}+3 H(\rho_{tot}+P_{tot})=  0$, where  ${\rho}_{tot} = \rho_m + \rho_\Lambda$  and $P_{tot}=P_\Lambda$.

Inspired by the discussions in Refs.~\cite{Dalal:2001dt,Guo:2007zk,Wei:2010uh},
the character of the $\Lambda(\alpha)$CDM models is given by
\begin{equation}
\label{eq:fa}
\frac{\rho_\Lambda}{\rho_m} = f(a),
\end{equation}
where $f(a)$ can be any function of the scale factor $a$.
For $f(a)\propto a^3$,   the coupling parameter $Q$ in Eqs.~(\ref{eq:eos1}) and (\ref{eq:eos2})
vanishes so that
 ${\rho_\Lambda}$ is a constant and  $\rho_m \propto a^{-3}$, representing the $\Lambda$CDM model.
From  Eqs.~(\ref{eq:friedmann1}) and (\ref{eq:fa}), we obtain
\begin{eqnarray}
\label{eq:12}
\Omega_\Lambda\equiv {8\pi G \rho_{\Lambda} \over 3 H^2}=\frac{f}{1+f},\\
\label{eq:12m}
\Omega_m \equiv {8\pi G \rho_{m} \over 3 H^2}
=\frac{1}{1+f}\,,
\end{eqnarray}
so that
 $\Omega_{m}+\Omega_{\Lambda}=1$.
Substituting Eq.~$\eqref{eq:fa}$  into Eq. $\eqref{eq:eos1}$ along with
Eqs.~(\ref{eq:12}) and (\ref{eq:12m}), we  get
\begin{equation}
\label{eq:Q}
Q=-H \rho_m \Omega_\Lambda\left(a\frac{f^\prime}{f}-3\right)=-H \rho_\Lambda \Omega_m \left(a\frac{f^\prime}{f}-3\right),
\end{equation}
where the prime ``$\prime$'' stands for a derivative with respect to $a$.
Subsequently,
from Eqs. $\eqref{eq:friedmann1}$ and $\eqref{eq:12}$
we
derive that
\begin{equation}
\label{eq:Delta}
\frac{\Delta \alpha}{\alpha}\equiv\frac{\alpha-\alpha_0}{\alpha_0}=\left(\frac{\Omega_\Lambda H^2}{H_0^2(1-\Omega_{m0})}\right)^{-1/6} -1,
\end{equation}
where the quantities with the subscript ``0'' correspond to those with $a=1$.
It is clear that, in the case of the $\Lambda$CDM model with
 $f \propto  \alpha^{3}$, ${\Delta \alpha}/{\alpha}=0$, implying  a constant ${\alpha}$.

 We now explore the possible forms for the $\Lambda(\alpha)$CDM models with a time-varying ${\alpha}$.
First of all,  $f( a )$ at present time with $a=1$ is given by
\begin{equation}
\label{eq:f0}
f_0 = f( a = 1 ) = \frac{\rho_{\Lambda 0}}{\rho_{m 0} }= \frac{1}{\Omega_{m 0}} -1.
\end{equation}
To simplify our discussions without loss of generality, we consider
\begin{equation}
\label{eq:f0a}
f(a)=f_0 a^{\xi(a)},
\end{equation}
where $\xi(a)$ is a function of $a$. For $\xi(a)=3$, it reduces to the $\Lambda$CDM model with the constant  $\alpha$.
Obviously, it is expected that
 $\xi(a)$ should be close to 3 so that the model does not deviate from the $\Lambda$CDM too much as required by
 the cosmological data.
  From Eqs.~(\ref{eq:12}) and (\ref{eq:12m}), we obtain
the explicit form
\begin{equation}
\label{eq:omegaE}
\frac{H^2}{H_0^2}=a^{-3}\left[\Omega_{m0}+(1-\Omega_{m0})a^{\xi(a)}\right]^{3/{\xi(a)}}.
\end{equation}
Consequently, we can derive that
\begin{eqnarray}
\label{eq:omega1}
\rho_{\Lambda}&=&\rho_{\Lambda 0} \frac{f}{(1+f)(1-\Omega_{m0})}\frac{H^2}{H_0^2}=\rho_{\Lambda 0} a^{\xi(a) -3}\left[(1-\Omega_{m0})a^{\xi(a)}+\Omega_{m0}\right]^{3/{\xi(a)}-1}\,,\\
\label{eq:omega2}
\rho_{m}&=&\frac{\rho_{\Lambda}}{f(a)}=\rho_{m 0}a^{ -3}\left[(1-\Omega_{m0})a^{\xi(a)}+\Omega_{m0}\right]^{3/{\xi(a)}-1}\,.
\end{eqnarray}

Similar to the  Chevallier-Polarski-Linder (CPL) EoS parameterization of $w = w_0 +w_a(1-a)$~\cite{CPL},
we take the simplest form for $\xi(a)$ to be CPL-like,  given by
\begin{equation}
\label{eq:xi(a)1}
\xi(a) = \xi_0 +\xi_1(1-a),
\end{equation}
with $\xi_0=3+u_0$. As $\xi(a)$ is close to 3,  $u_0\to 0$.
The function for $\xi(a)$ in Eq.~(\ref{eq:xi(a)1}) is labelled as $\Lambda(\alpha)$CDM1.
Note that  this $\Lambda(\alpha)$CDM model along with the special case with $\xi_1 = 0$
has been discussed in Ref.~\cite{Wei:2016moy}.
The relation of $\xi_1$ and ${\Delta \alpha}/{\alpha}$ can be explicitly written as
\begin{eqnarray}
\label{eq:xi-alpha}
\xi_1 =\frac{\ln \left(a^{\xi_0}(f_0+1)({\Delta \alpha\over\alpha}+1)^6\frac{H^2}{H_0^2} -f_0a^{\xi_0} \right) }{(a-1)\ln a}\,,
\end{eqnarray}
It is easy to check that if $\Delta \alpha/\alpha$=0, the model becomes $\Lambda$CDM with $\xi_0=3$ and $\xi_1=0$.

To illustrate the feature of $\xi(a)$,
we also consider  $u_0 = 0$
 in Eq.~(\ref{eq:xi(a)1}), and refer to
the resulting function of
\begin{equation}
\label{eq:xi(a)2}
{\xi(a)}=3+\xi_1(1-a),
\end{equation}
 as  $\Lambda(\alpha)$CDM2, which was not studied  in Ref.~\cite{Wei:2016moy}.
In the following, we will concentrate on the two models of $\Lambda(\alpha)$CDM1 and $\Lambda(\alpha)$CDM2.

\subsection{Linear perturbation theory}
\label{sec:P}
In order to consider whether the models  can be established in the dynamical universe,
the linear perturbation theory should be
taken into account to examine the dynamics of the
$\Lambda(\alpha)$CDM models.
Here, we will study the growth equations of the density perturbation for the models
based on the standard linear perturbation theory~\cite{Ma:1995ey}.
In the synchronous gauge, the metric  is given by
\begin{eqnarray}
\label{eq:Metric}
ds^{2}=a^{2}(\tau)[-d\tau^{2}+(\delta_{ij}+h_{ij})dx^{i}dx^{j}],
\end{eqnarray}
where  $i,j=1,2,3$, $\tau$  is the conformal time, and
\begin{eqnarray}
h_{ij}= \int d^3 k e^{i\vec k\vec x}\left[\hat{k}_i \hat{k}_j h(\vec k,\tau)+6\left(\hat{k}_i\hat{k}_j-\frac{1}{3}\delta_{ij}\right)\eta(\vec k,\tau)\right]\,,
\end{eqnarray}
with  the $k$-space unit vector of $\hat{k}= \vec k/k$   and two scalar perturbations of $h(\vec k,\tau)$ and $\eta(\vec k,\tau)$.
From the conservation equation of
$\nabla^{\nu}( T^{m}_{\mu\nu}+T^{\Lambda}_{\mu\nu})=0$ with $\delta T^{0}_{0}=\delta \rho_{m}$, $\delta T^{0}_{i}=-T^{i}_{0}=(\rho_{m}+P_{m})v^i_m$ and $\delta T^{i}_{j}=\delta P_{m}\delta ^{i}_{j}$.

As explicitly shown in Refs.~\cite{DEP2,Grande:2008re}, there are two basic perturbation equations, given by
 \begin{eqnarray}
\label{eq:theta1}
\sum_{i = \Lambda, m} \delta \rho_i+3\delta(\frac{H}{a})(\rho_i+P_i)+3\frac{H}{a}(\delta\rho_i+\delta P_i) = 0,
\end{eqnarray}
\begin{eqnarray}
\label{eq:theta2}
\sum_{i = \Lambda, m}\dot\theta_i(\rho_i+P_i)+\theta_i (\dot\rho_i+\dot P_i+5H(\rho_i+P_i)) = \frac{k^2}{a}\sum_{i = \Lambda, m}\delta P_i,
\end{eqnarray}
where $H=da/(a d\tau)$ in terms of conformal time $\tau$, $\delta \rho_i$ represent the density fluctuations,
and $\theta_i$ are the corresponding velocities. As there is no  peculiar velocity for dark energy, we take $\theta_\Lambda=0$.
In addition, we assume that $\delta\rho_m\gg \delta\rho_\Lambda$ and $\delta\dot{\rho}_m \gg \delta\dot{\rho}_\Lambda$  in our models.
From Eqs.~(\ref{eq:theta2}), we get
\begin{eqnarray}
{\dot{\theta}_m}+{\theta}_m({2H}-\frac{\dot{\rho}_{\Lambda}}{ {\rho}_m})=-\frac{k^2}{a}\frac{\delta \rho_{\Lambda}}{\rho_m},
\end{eqnarray}
resulting in the momentum conservation equation in this gauge, given by
\begin{eqnarray}
\dot{v}_m+H v_m+v_m\frac{\dot{\rho}_{\Lambda}}{ {\rho}_m}=\frac{\delta \rho_{\Lambda}}{\rho_m},
\end{eqnarray}
based on $\theta_m=-{k^2}v_m/{a}+{\cal O} (2)$ with ${\cal O}(2)$ referring  to the second order perturbations.
Due to the remaining gauge freedom left~\cite{Wang:2014xca,Wang:2013qy} in
the synchronous gauge, one can take
the zero velocity of matter, $i.e.$, $v_m=0$, which leads to  $\delta \rho_{\Lambda}=\theta_m=0$.
As a result,  in our calculations we can choose that
$\delta\rho_\Lambda\to 0$ and $\theta_m\to 0$. For the matter perturbation,
the growth equations are given by
\begin{eqnarray}
\label{eq:pert1}
&& \dot{\delta}_m=-(1+w_m)\left(\theta_{m}+\frac{\dot{h}}{2}\right)-3H\left(\frac{\delta P_{m}}{\delta \rho_{m}}-w_{m}\right)\delta_{m}-\frac{Q}{\rho_{m}}\delta_{m} \,, \\
\label{eq:pert2}
&& \dot{\theta}_m=-H(1-3w_{m})\theta_{m}-\frac{\dot{w}_{m}}{1+w_{m}}\theta_{m}+\frac{\delta P_{m}/\delta \rho_{m}}{1+w_m}\frac{k^{2}}{a^{2}}\delta_m-\frac{Q}{\rho_{m}}\theta_{m} \,,
\end{eqnarray}
where $\delta_i\equiv \delta \rho_i/\rho_i$
and $Q$ is the coupling term in Eqs.~(\ref{eq:eos1}) and (\ref{eq:eos2}).
 To simplify  our calculations in Eqs.~(\ref{eq:pert1}) and (\ref{eq:pert2}),
we will take $\delta P_m/\delta\rho_m = w_m=\dot{w}_{m}=0$.


\section{Numerical calculations }
\label{sec:observation}
 We use {\bf CAMB} and {\bf CosmoMC} to do the numerical calculations for the  two  models of $\Lambda(\alpha)$CDM1 and $\Lambda(\alpha)$CDM2.
 We fit the model parameters in Eqs.~(\ref{eq:xi(a)1}) and (\ref{eq:xi(a)2}) with
 the observational data by the {\bf MCMC} method.
  In the calculation, we need to modify the {\bf CAMB} program with
 Eqs.~(\ref{eq:pert1}) and (\ref{eq:pert2}) given by the linear perturbation for the models.
 In order to have  more accurate results, we take the datasets, which contain
  the CMB temperature fluctuations from
   Planck 2015 with TT, low-l polarizations
  and CMB lensing from SMICA~\cite{Adam:2015wua,Aghanim:2015xee,Ade:2015zua},
  the BAO data from 6dF Galaxy Survey~\cite{Beutler:2011hx} and BOSS~\cite{Anderson:2013zyy},
and  the Type Ia supernovae data from Supernova Legacy Survey~\cite{Bazin:2011xw}.
In addition, we include 313 data points of ${\Delta \alpha}/{\alpha}$ from
the absorption systems in the spectra of distant quasars with 0.2223 $\leqslant z_{abs}\leqslant$
4.1798 in the analysis. Note that among these data,
293 were published in 2012~\cite{King:2012id}\footnote{Although there are
141 and 154 quasar absorption systems from the Keck
Observatory in Hawaii and  Very Large Telescope (VLT) in Chile, respectively,
two outliers with J194454+770552 at $z_{abs} = 2.8433$ and J000448-415728 at $z_{abs} = 1.5419$
have been excluded in Refs.~\cite{King:2012id,30,31,32}}, while  20 of them are the new ones~\cite{Wilczynska:2015una}.
It is interesting to emphasize that  ${\Delta \alpha}/{\alpha}=\mathcal{O}(10^{-5})$ for  all  data points.
The $\chi^2$ value is
 given by
 \begin{eqnarray}
\label{eq:chi}
{\chi^2}={\chi^2_{CMB}}+{\chi^2_{BAO}}+{\chi^2_{WL}}+{\chi^2_{SN}}+{\chi^2_\alpha} \,,
\end{eqnarray}
where $\chi^2_j$ ($j=CMB,BAO,WL,SN)$ are the $\chi^2$ standard calculations and $\chi^2_\alpha$ is given by
\begin{eqnarray}
 {\chi^2_\alpha}=\sum_{i} \frac{[{\Delta \alpha/\alpha}_{th,i}-{\Delta \alpha/\alpha}_{obs,i}]^2}{\sigma ^2_i} \,,
\label{eq:chi2}
\end{eqnarray}
with $\sigma^2_i=\sigma^2_{stat,i}+\sigma^2_{rand,i}$, defined in Refs.~\cite{King:2012id,30,31,32, Wilczynska:2015una}.

\begin{figure}
\centering
\includegraphics[width=0.96 \linewidth]{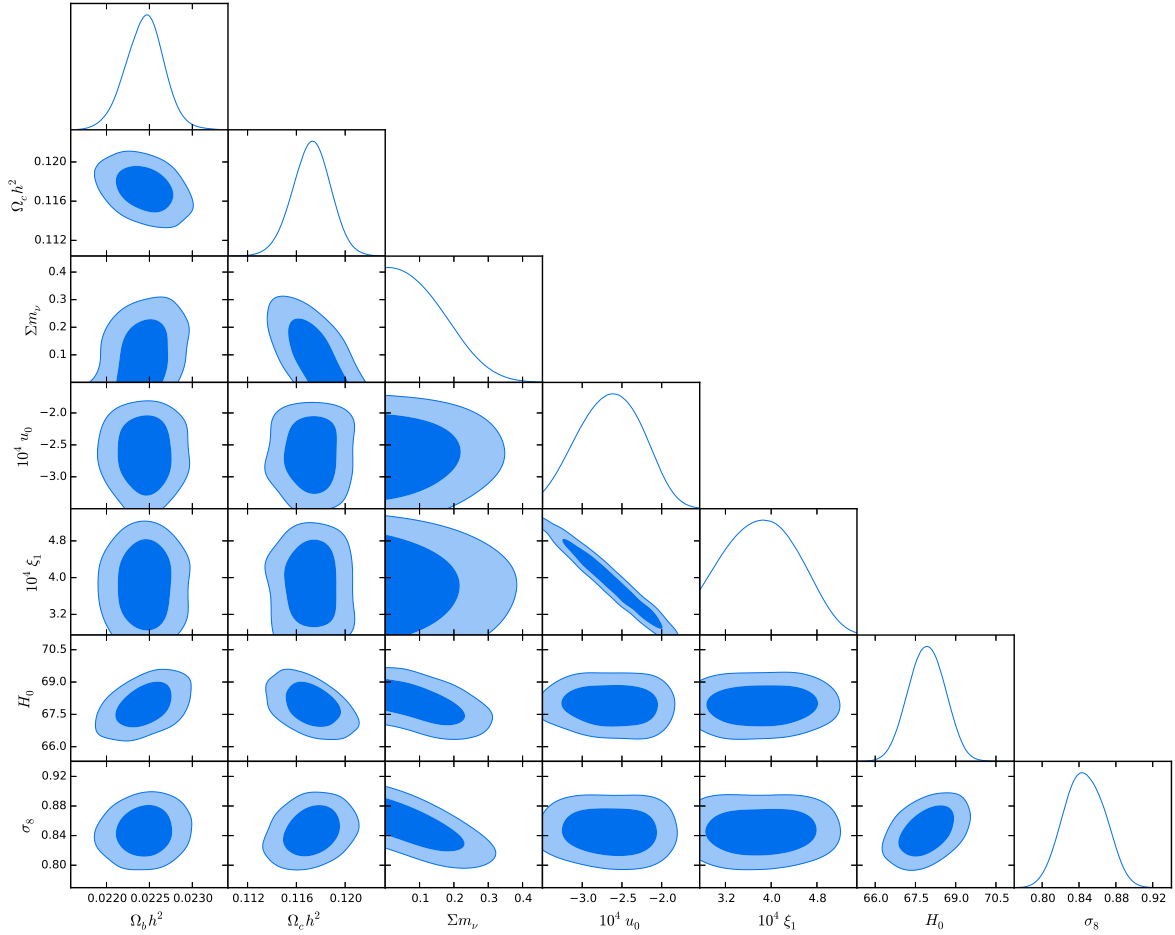}
\caption{One and two-dimensional distributions of $\Omega_b h^2$, $\Omega_c h^2$, $\sum m_\nu$, $10^4u_0$, $10^4\xi_1$, $H_0$,
and $\sigma_8$ for $\Lambda(\alpha)$CDM1 with ${\xi(a)}=3+u_0+\xi_1(1-a)$,
where the contour lines represent 68$\%$~ and 95$\%$~ C.L., respectively.}
\label{fg:1}
\end{figure}

\begin{figure}
\centering
\includegraphics[width=0.96 \linewidth]{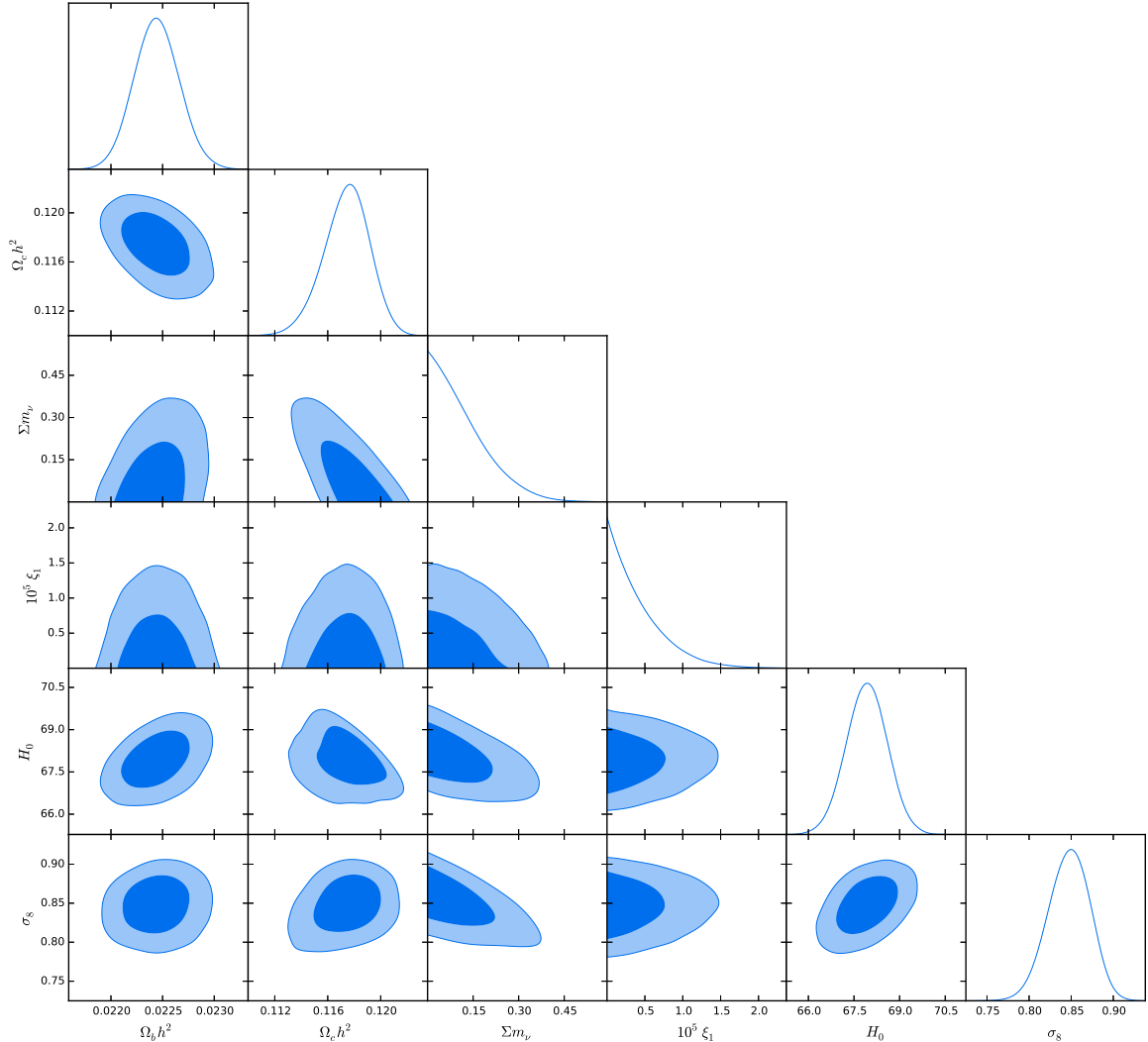}
\caption{Legend is the same as Fig.~\ref{fg:1} but for $\Lambda(\alpha)$CDM2 with ${\xi(a)}=3+ \xi_1(1-a)$. }
\label{fg:2}
\end{figure}

In Table.~\ref{tab:1}, we list the priors for cosmological parameters with the models in Eqs.~(\ref{eq:xi(a)1}) and (\ref{eq:xi(a)2}).
\begin{table}[ht]
\begin{center}
\caption{Priors for cosmological parameters with  $\Lambda(\alpha)$CDM1: ${\xi(a)}=3+u_0+\xi_1(1-a)$
and  $\Lambda(\alpha)$CDM2: ${\xi(a)}=3+\xi_1(1-a)$  }
\begin{tabular}{@{}c|c@{}} \toprule
Parameter & Prior
\\ \hline
 $u_0$ in $\Lambda(\alpha)$CDM1 & $-3.5 \times 10^{-4}\leq u_0 \leq -1.5 \times 10^{-4}$
\\ \hline
 $\xi_1$ in $\Lambda(\alpha)$CDM1 & $2.75 \times 10^{-4} \leq \xi_1 \leq 5.5 \times 10^{-4}$
\\ \hline
 $\xi_1$ in $\Lambda(\alpha)$CDM2 & $0 \leq \xi_1 \leq 5 \times 10^{-5}$
\\ \hline
Baryon density & $0.5 \leq 100\Omega_bh^2 \leq 10$
\\ \hline
CDM density & $10^{-3} \leq \Omega_ch^2 \leq 0.99$
\\ \hline
Optical depth & $0.01 \leq \tau \leq 0.8$
\\ \hline
Neutrino mass sum& $0 \leq \Sigma m_{\nu} \leq 2$~eV
\\ \hline
$\frac{\mathrm{Sound \ horizon}}{\mathrm{Angular \ diameter \ distance}}$  & $0.5 \leq 100 \theta_{MC} \leq 10$
\\ \hline
Scalar power spectrum amplitude & $2 \leq \ln \left( 10^{10} A_s \right) \leq 4$
\\ \hline
Spectral index & $0.8 \leq n_s \leq 1.2$
\\\botrule
\end{tabular}
\label{tab:1}
\end{center}
\end{table}
In Fig.~\ref{fg:1}, we present our global fit  from various datasets for
$\Lambda(\alpha)$CDM1 with ${\xi(a)}=3+u_0+\xi_1(1-a)$, where the values of $\sigma_8$ are given at $z = 0$.
Similarly, in Fig.~\ref{fg:2} we show our results for $\Lambda(\alpha)$CDM2 with ${\xi(a)}=3+ \xi_1(1-a)$.

We summarize our fitting results for the two $\Lambda(\alpha)$CDM models in Table~\ref{tab_1DM},
 in which we also include those in $\Lambda$CDM.
 It is clear that the model parameters of $u_0$ and $\xi_1$ in the $\Lambda(\alpha)$CDM
models are zero in the limit of $\Lambda$CDM.
\begin{table}[ht]
\caption{Summary of the fitting results for $\Lambda(\alpha)$CDM1 with ${\xi(a)}=3+u_0+\xi_1(1-a)$ and $\Lambda(\alpha)$CDM2 with
${\xi(a)}=3+\xi_1(1-a)$ as well as those for  $\Lambda$CDM, where  313
$\Delta \alpha/\alpha$ data are used the limits are given at 68$\%$ C.L.}
{\begin{tabular}{@{}c|c|c|c|c|c|c|c|c@{}} \toprule
 Model &{ $100 \Omega_bh^2$}&{ $100 \Omega_ch^2$}&{ $H_0$} &{ $\sigma_8$}&{ ${\Sigma m_\nu \over \mathrm{eV}}$ } &{ $10^4u_0$} &{ $10^4\xi_1$} &{ $\chi^2_{best-fit}$} \\
\colrule
{ $\Lambda(\alpha)$CDM1 } & $ 2.24 \pm 0.02$  & $11.7 ^{+0.16}_{-0.14}$ & $67.90\pm 0.69$ & $ 0.844^{+0.027}_{-0.023}$ &
$0.123^{+0.032}_{-0.122}$&$-2.67^{+0.39}_{-0.40}$&$3.95^{+0.59}_{-0.64}$&$1844.132$\\
\hline
{ $\Lambda(\alpha)$CDM2 } & $ 2.24 \pm 0.02$  & $11.7 ^{+0.19}_{-0.15}$ & $67.91^{+0.71}_{-0.68}$ & $ 0.847^{+0.028}_{-0.025}$ &$<0.144$&$-$&$0.0409^{+0.0080}_{-0.0409}$&$1870.837$\\
\hline
{ $\Lambda$CDM } & $ 2.25 \pm 0.02$  & $11.7 ^{+0.20}_{-0.16}$ & $67.83^{+0.74}_{-0.67}$ & $ 0.843^{+0.027}_{-0.024}$ &$<0.165$&$0$&$0$&$1869.854$\\
\botrule
\end{tabular}\label{tab_1DM}}
\end{table}

From the table, we find that $u_0=(-2.667^{+0.393}_{-0.398})\times 10^{-4} $
and $\xi_1=(3.953^{+0.591}_{-0.641})\times 10^{-4}$
in ${\xi(a)}=3+u_0+\xi_1(1-a)$ of  $\Lambda(\alpha)$CDM1 and $0.409^{+0.080}_{-0.409}\times 10^{-5}$
in  ${\xi(a)}=3+\xi_1(1-a)$ of  $\Lambda(\alpha)$CDM2
with the best fitted $\chi^2$ values being $1844.132$
 and  $1870.837$, respectively.
  As expected, the two-parameter  model of $\Lambda(\alpha)$CDM1 gives the lowest value of $\chi^2_{best fit}$, while
the one-parameter one of $\Lambda(\alpha)$CDM1 leads to a slightly larger $\chi^2_{best fit}$ than $\Lambda$CDM.
The lower bound of $\xi_1$ in $\Lambda(\alpha)$CDM2 is due to its prior set from zero
 in Table~\ref{tab:1}. Without such a prior, a negative value at $O(10^{-5})$ for $\xi_1$ is also possible.
 It is clear that our fitting results for $\Lambda(\alpha)$CDM1 with two free model parameters are better than those for
 $\Lambda(\alpha)$CDM2 with a single one.
  Comparing with the best-fit values  $u_0 $ and $\xi_1$
   in $\Lambda(\alpha)$CDM1  given by Ref.~\cite{Wei:2016moy},
  our results are slightly different due to
   the different fitting method and cosmological data in our calculations.

   From Table~\ref{tab_1DM},
   it is interesting to see  that our fitting result for ${\Sigma m_\nu}$ in $\Lambda(\alpha)$CDM1
   is $0.123^{+0.032}_{-0.122}$ eV at $68\%$ C.L.,
    whereas  that for $\Lambda(\alpha)$CDM2 only gives
the upper bound of $0.152$  eV.
   We note that the values of $\sigma_8$ in our two models are both slightly larger that that in   $\Lambda$CDM.

\begin{figure}[!t]
\centering
\includegraphics[angle=-90,width=75mm]{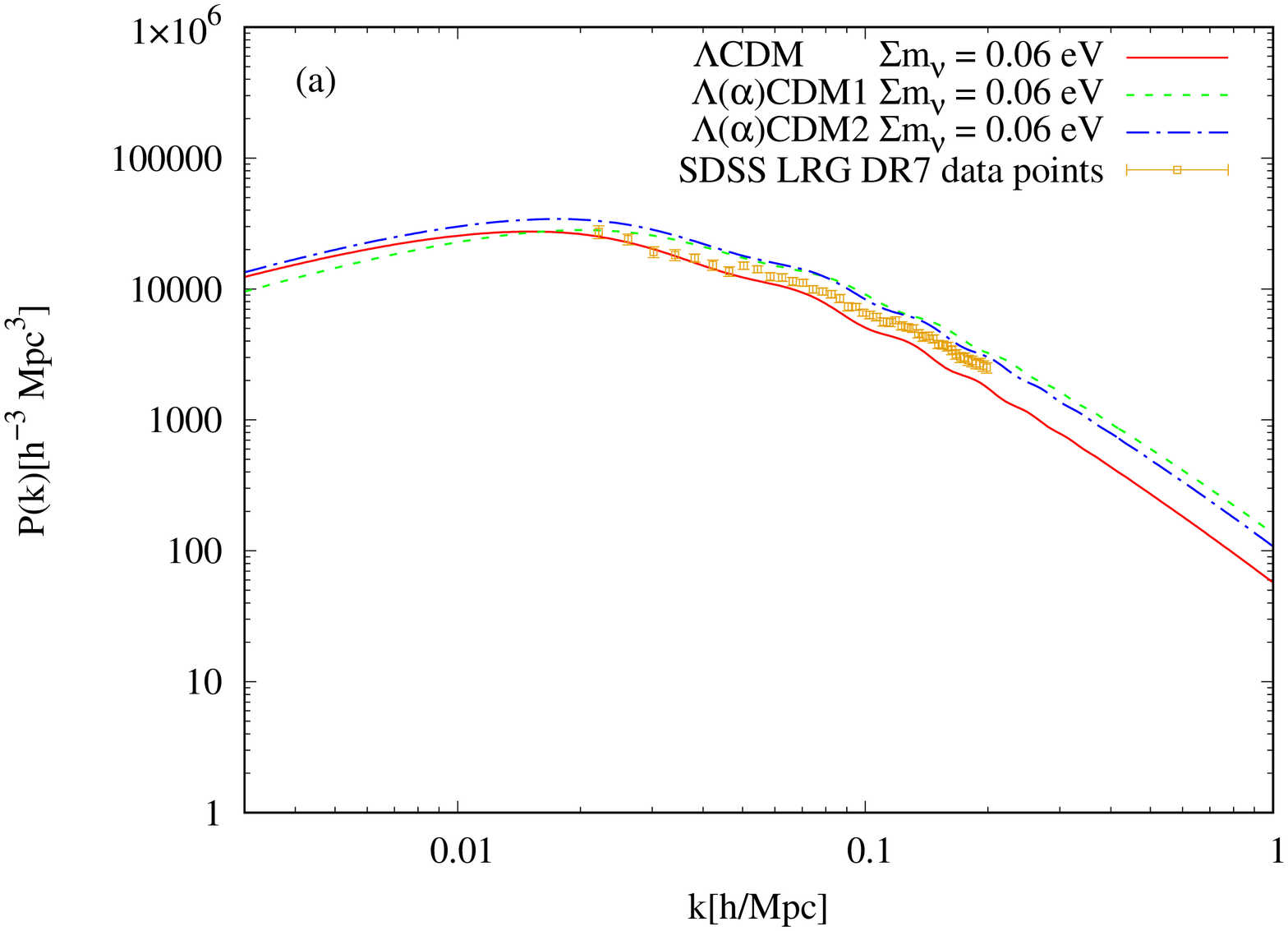}
\includegraphics[angle=-90,width=75mm]{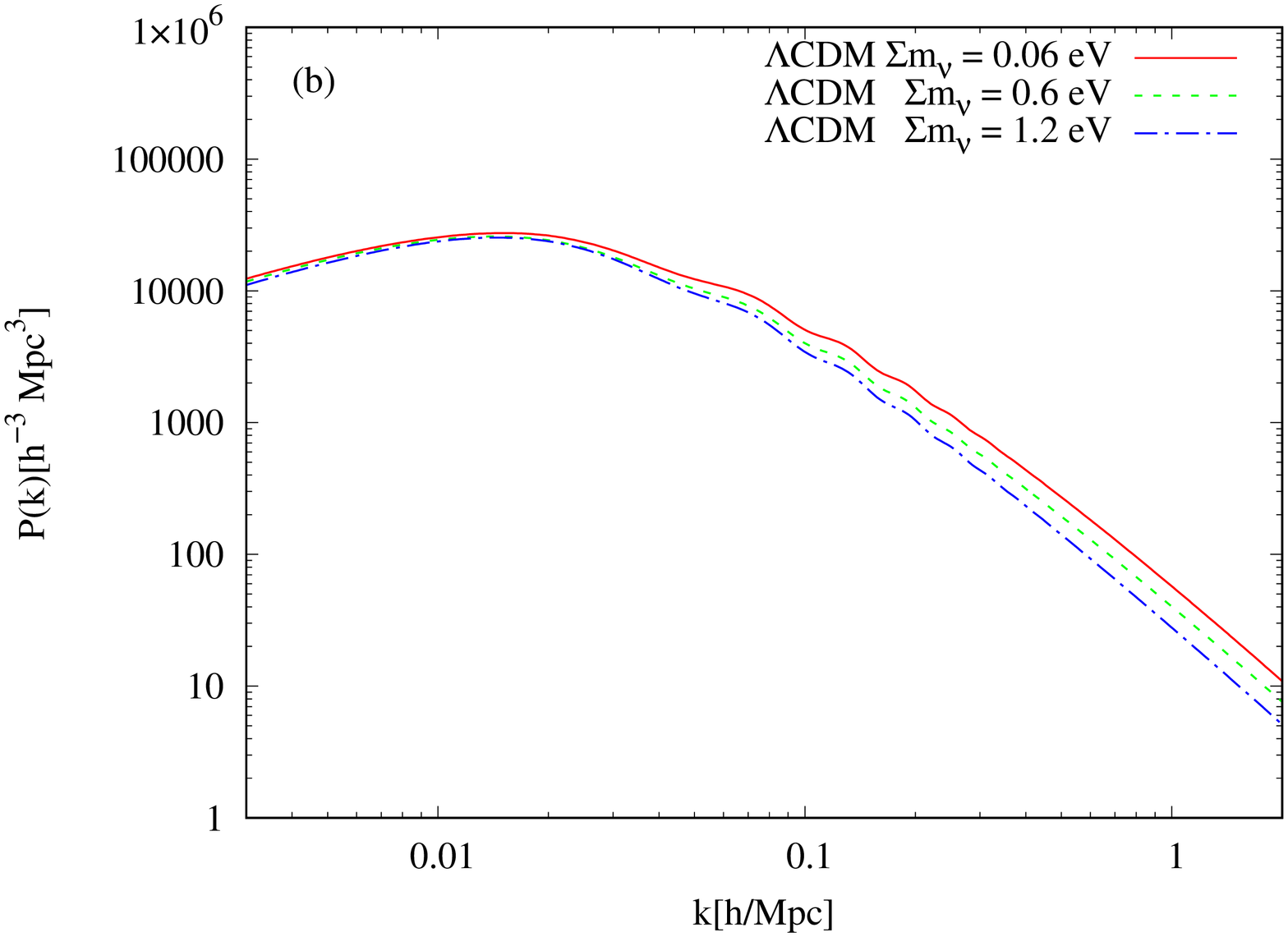}
\\
\includegraphics[angle=-90,width=75mm]{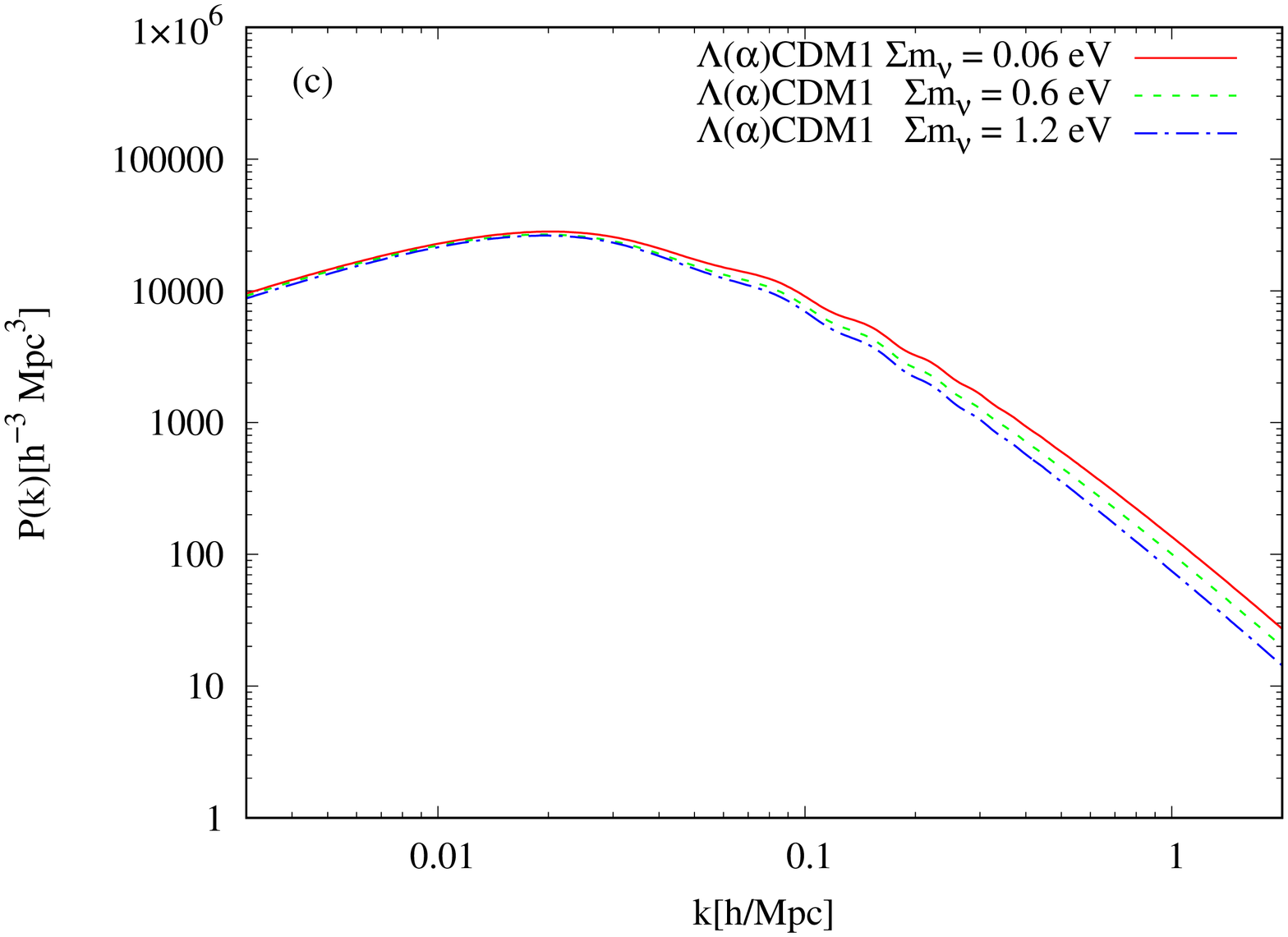}
\includegraphics[angle=-90,width=75mm]{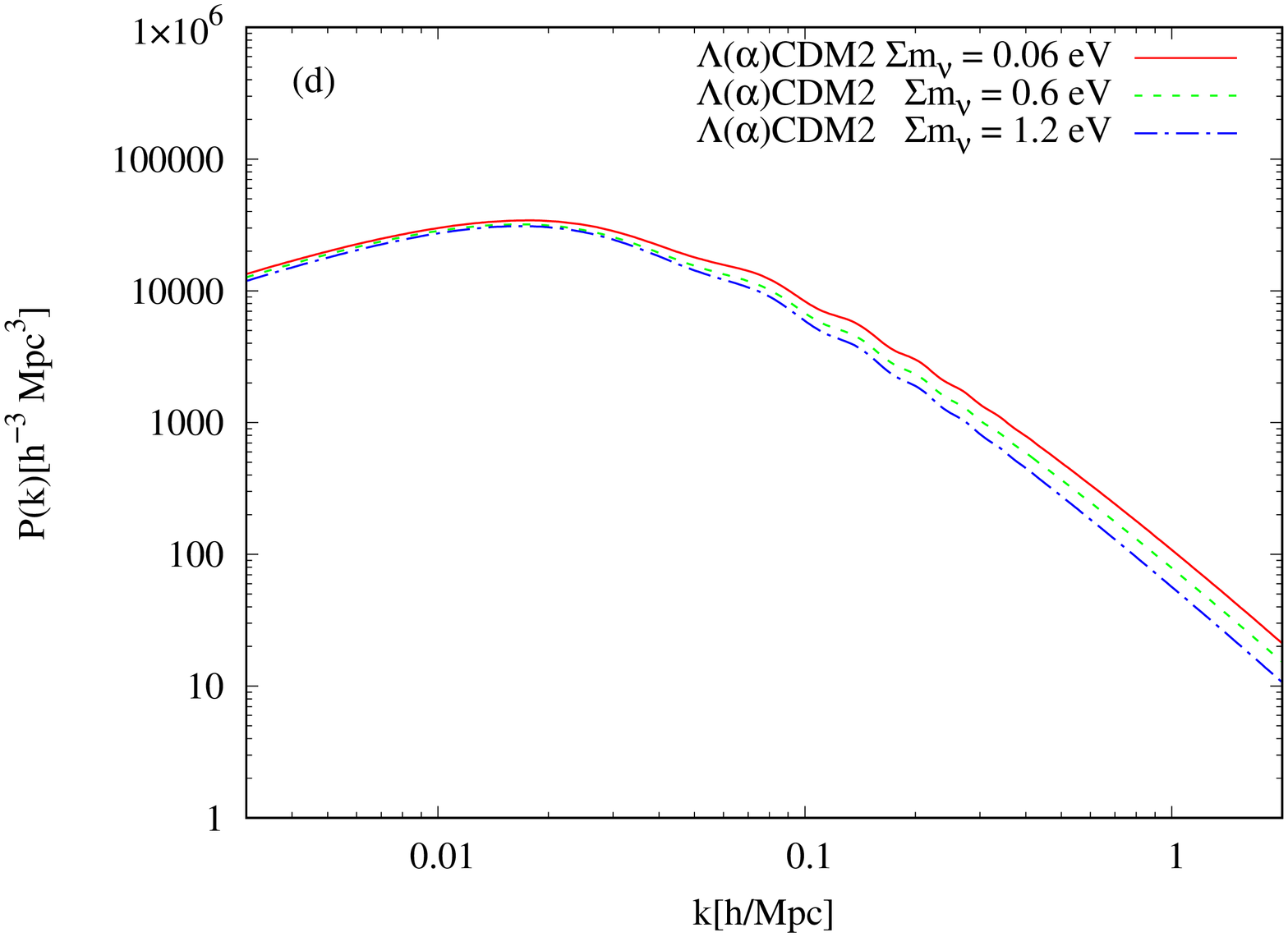}
\caption{Matter power spectra of $P(k)$ as  functions of the wavelength $k = 2\pi/\lambda$
in (a) $\Lambda(\alpha)$CDM and $\Lambda$CDM with $\Sigma m_\nu$=0.06 eV along with the observational data,
(b) $\Lambda$CDM with $\Sigma m_\nu$=0.06, 0.6 and 1.2 eV,
(c) $\Lambda(\alpha)$CDM1 with $\Sigma m_\nu$=0.06, 0.6 and 1.2 eV,
and
(d)  $\Lambda(\alpha)$CDM2 with $\Sigma m_\nu$=0.06, 0.6 and 1.2 eV.}
\label{fg:3}
\end{figure}

\begin{figure}
\centering
\includegraphics[width=0.96 \linewidth]{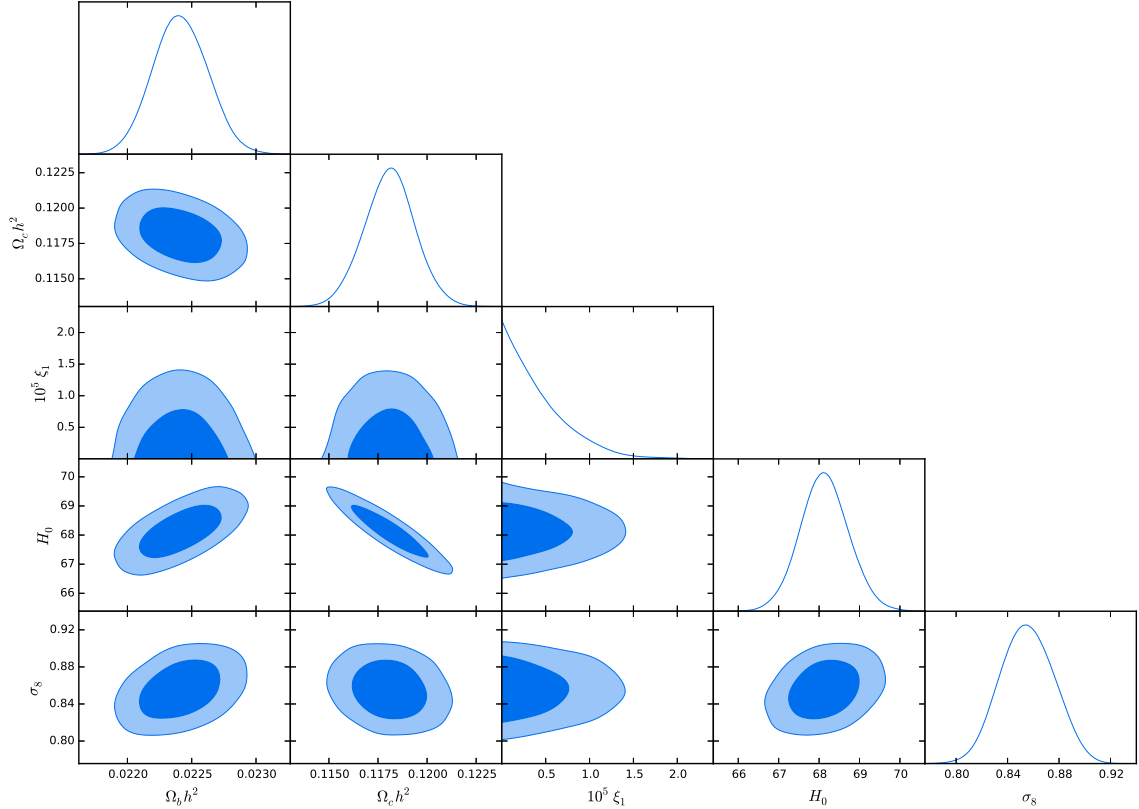}
\caption{One and two-dimensional distributions of $\Omega_b h^2$, $\Omega_c h^2$, $10^5\xi_1$, $H_0$,
and $\sigma_8$ for $\Lambda(\alpha)$CDM2 with ${\xi(a)}=3+ \xi_1(1-a)$ with $\Sigma m_\nu$=0.06 eV,
where the contour lines represent 68\% and 95\% C.L., respectively. }
\label{Newfig}
\end{figure}

        To understand the behaviors of ${\Sigma m_\nu}$ in the various models,
        we show the matter power spectra as functions of the wavelength $k=2\pi/\lambda$
 in the $\Lambda$CDM as well as  $\Lambda(\alpha)$CDM1 and
$\Lambda(\alpha)$CDM2 models in Fig.~\ref{fg:3}.
To exhibit the trend of the matter power spectrum  in terms of $\Sigma m_\nu$, we present
Figs.~\ref{fg:3}b, \ref{fg:3}c and \ref{fg:3}d for $\Lambda$CDM,  $\Lambda(\alpha)$CDM1 and
$\Lambda(\alpha)$CDM2 with $\Sigma m_\nu$=0.06, 0.6 and 1.2 eV, respectively.
From Fig.~\ref{fg:3}a with the fixed value of $\Sigma m_\nu=0.06$ eV,
we find that, in comparison with $\Lambda$CDM,
the matter power spectrum in $\Lambda(\alpha)$CDM1(2) gets enhanced for the most (all) region of $k$,
whereas that in $\Lambda(\alpha)$CDM1 slightly suppressed for the low values of $k$.
On the other hand,  the value of  $\Sigma m_\nu$ increases the suppression factor for the
matter power spectrum within  the same model as illustrated in Figs.~\ref{fg:3}b-d.
The enhancement behaviors of the matter power spectra in $\Lambda(\alpha)$CDM are similar
to the cases in the viable $f(R)$  gravity models as studied in Ref.~\cite{f(R)NuMass}.
In Fig.~\ref{Newfig},
we depict
 our results of $\Lambda(\alpha)$CDM2 for chains with $\Sigma m_\nu$ fixed to be 0.06 eV to illustrate
the best-fit parameters.
We also summary our fit for $\Lambda(\alpha)$CDM2 $\Sigma m_\nu=0.06$ eV in Table~\ref{Newtab}, where the corresponding results
for $\Lambda$CDM are also given.
\begin{table}[ht]
\caption{Summary of the fitting results for $\Lambda(\alpha)$CDM2 with
${\xi(a)}=3+\xi_1(1-a)$ and ${\Sigma m_\nu =0.06 \mathrm{eV}}$,  the limits are given at 68$\%$ C.L.}
{\begin{tabular}{@{}c|c|c|c|c|c|c@{}} \toprule
 Model &{ $100 \Omega_bh^2$}&{ $100 \Omega_ch^2$}&{ $H_0$} &{ $\sigma_8$} &{ $10^5\xi_1$} &{ $\chi^2_{best-fit}$} \\
\colrule
{ $\Lambda(\alpha)$CDM2 } & $ 2.24 \pm 0.02$  & $11.8 \pm 0.1$ & $68.11 ^{+0.59}_{-0.60}$ & $ 0.853\pm 0.021$ &$0.418^{+0.089}_{-0.418}$&$1872.230$\\
\hline
{ $\Lambda$CDM } & $ 2.24 \pm 0.02$  & $11.8 \pm 0.1$ & $68.13 ^{+0.59}_{-0.60}$ & $ 0.855 ^{+0.019}_{-0.022}$ &$0$&$1870.574$\\
\botrule
\end{tabular}\label{Newtab}}
\end{table}
Note that we are able to get a good fit for $\Lambda(\alpha)$CDM1 with $\Sigma m_\nu\sim 0$ as it favors a large $\Sigma m_\nu$
as indicated in Table~\ref{tab_1DM}.
It is interesting to see that the best-fit value of $\chi^2$ for $\Lambda(\alpha)$CDM2
with  $m_\nu$ fixed to be 0.06 eV is 1872.230,  which is  larger than 1870.837 without fixing $m_\nu$.
Note that the corresponding values of $\chi^2_{best-fit}$ are 1870.574 and 1869.854 for $\Lambda$CDM with and without fixed $m_\nu$, respectively.

Finally, we remark that the model parameter of $\xi_1$ can be also  constrained directly by using
Eq.~(\ref{eq:xi-alpha}). For example, one can show that $\xi_1$ in $\Lambda(\alpha)$CDM is $O(10^{-4})$
for ${\Delta \alpha}/{\alpha}=\mathcal{O}(10^{-5})$.

\section{Conclusions}
\label{sec:CONCLUSIONS}

We have studied the $\Lambda(\alpha)$CDM models with $\Lambda(\alpha)\propto\alpha^{-6}$, in which the fine structure constant
$\alpha$ varies in time with the data of ${\Delta \alpha}/{\alpha}=\mathcal{O}(10^{-5})$.
In particular, we have concentrated on two specific $\Lambda(\alpha)$CDM models in Eqs.~(\ref{eq:xi(a)1}) and (\ref{eq:xi(a)2}) with two and one model parameters, respectively. We have performed global fits on the two models by using the available cosmological data in the {\bf CAMB} and {\bf CosmoMC} packages together with 313 data points for ${\Delta \alpha}/{\alpha}$ from distant quasars.
We have shown that the model parameters are constrained to be around $10^{-4}$, which are similar to those given by
Ref.~~\cite{Wei:2016moy} but with  more accurate outcomes.
For $\Lambda(\alpha)$CDM1, we have derived an interesting fitting value of ${\Sigma m_\nu}$ is $0.123^{+0.032}_{-0.122}$eV,
which gives not only an upper bound of 0.155 eV but a lower one of $9.87\times 10^{-4}$ eV,
instead  of  the only upper bounds  in most of cosmological models, including $\Lambda$CDM and $\Lambda(\alpha)$CDM2.
 In addition,  we have found that the best fitted $\chi^2$ values are 1844.132 and 1870.837
for the two models of  $\Lambda(\alpha)$CDM1 and $\Lambda(\alpha)$CDM2, respectively.

\section*{Acknowledgments}

We thank Dr. Joan Sola, Dr. Chung-Chi Lee, Dr. Ling-Wei Luo, Dr. Emmanuel N. Saridakis and Dr. Hao Wei
for the useful discussions.
This work was supported in part by National Center for Theoretical Sciences,
MoST (MoST-104-2112-M-007-003-MY3) and NSFC (11547008).

\end{document}